# Recombination or Transport? Understanding Electronic and Ionic Processes in Perovskite Solar Cells by Combining Impedance and Light-Modulation Spectroscopies

Agustin O. Alvarez,[a]* Athira Shaji,[a] Max Mundt,[a] Biruk A. Seid,[b] Riccardo Pallotta,[c] Elena Mas-Marzá,[d] Giulia Grancini,[c] Felix Lang,[b] Francisco Fabregat-Santiago,[d]* and Bruno Ehrler[a]*

[a] LMPV - Sustainable Energy Materials Department, AMOLF, 1098 XG Amsterdam, The Netherland

[b] ROSI Freigeist Juniorgroup, Institut für Physik und Astronomie, University of Potsdam, 14476 PotsdamGolm, Germany

[c] Department of Chemistry, University of Pavia, Via T. Taramelli 14, 27100 Pavia, Italy

[d] Institute of Advanced Materials, Universitat Jaume I, 12006 Castelló de la Plana, Spain

E-mail: a.alvarez@amolf.nl; fabresan@uji.es; b.ehrler@amolf.nl

## TOC

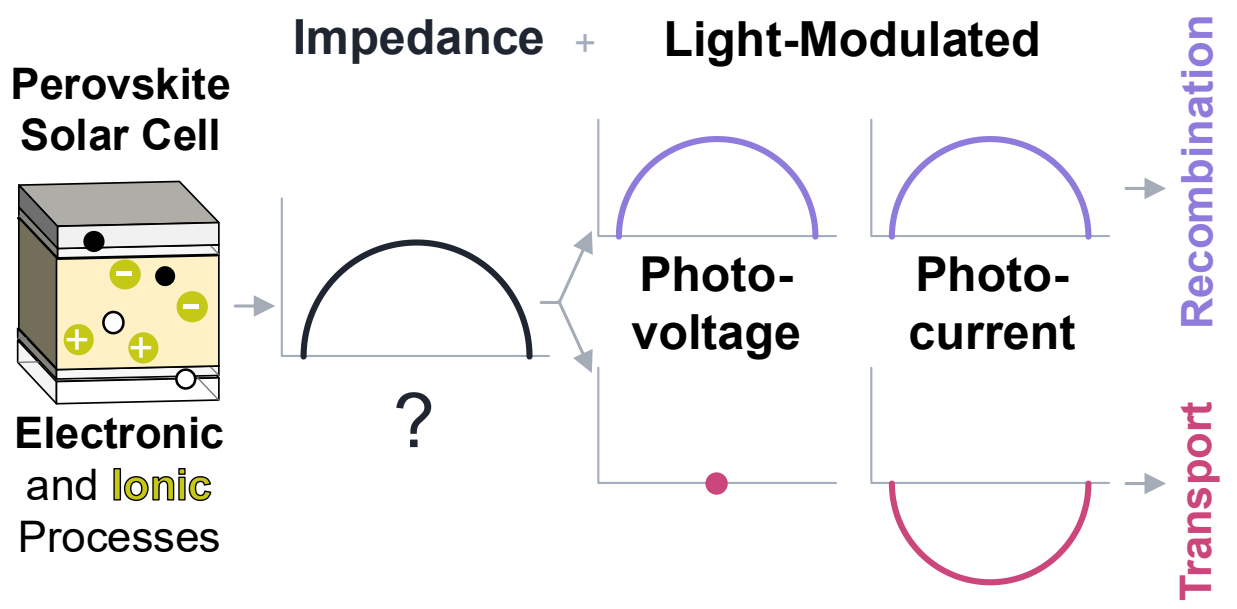

Different electronic and ionic mechanisms in perovskite solar cells can produce similar impedance spectra. Light-modulated photovoltage and photocurrent measurements reveal their distinct effects as recombination-type and transport limitations.

## Abstract

Impedance spectroscopy (IS) is widely used to investigate dynamic processes in perovskite solar cells (PSCs), but impedance alone cannot distinguish whether a process, observed as an arc in the Nyquist plot, is associated with recombination or transport. Here, we combine IS with intensity-modulated photocurrent (IMPS) and photovoltage (IMVS) spectroscopies, which modulate photogeneration, and show directly how each process changes the photocurrent and photovoltage. We apply this methodology to three different PSCs. In a PSC with a high-power conversion efficiency of 24.3%, we identify two processes (two IS arcs) as recombination: a fast electronic process and a slow ion-related process. In a PSC with non-optimized contacts, we observe the same two processes and an additional process, that we identify as a transport limitation because it produces no change in photovoltage and reduces the photocurrent. In a lower-efficiency PSC, the ion-related process, showing an inverted (inductive) IS arc, instead increases recombination. For each device, a simple equivalent circuit that explicitly includes the photogenerated current fits the three techniques simultaneously. These results show that combining IS with IMVS and IMPS resolves the ambiguity in interpreting impedance spectra and clarifies the functional roles of electronic and ionic processes in PSCs.

## 1. Introduction

Metal halide perovskites are unusual among photovoltaic materials in that they behave as mixed ionic-electronic conductors: fast electronic processes and comparatively slow ionic migration coexist within the same absorber and interact with one another[1, 2]. This coupling manifests itself in many different ways and has been probed with a range of techniques, including current-voltage hysteresis [3], transmission electron microscopy (TEM) [4], photoluminescence [5], and time-of-flight secondary ion mass spectrometry (ToF-SIMS) [6]. Ion migration is typically associated with limited reproducibility and poor long-term stability of perovskite solar cells (PSCs), which are among the central obstacles to their commercialization [2]. The net effect of ion migration on device performance can be either detrimental or beneficial [7-10]. Moreover, the same ionic-electronic coupling has been exploited for halide-perovskite memristive and neuromorphic devices [11, 12]. Properly identifying and characterizing the processes that govern the mixed ionic-electronic conduction is therefore essential.

Impedance spectroscopy (IS), where devices are probed with a modulation of the voltage at varying frequencies and the current response is measured (or vice versa), is the most widely used technique to probe the internal processes in PSCs. Each process that corresponds to a characteristic timescale, for example, charge recombination or charge extraction, is resolved as an arc in the Nyquist plot [13]. Perovskite devices show a few typical responses. In efficient devices, the most common response is two arcs, with a high-frequency arc usually attributed to electronic recombination and a low-frequency arc often attributed to ion-related processes [14-16]. Whether the latter is related to an ion-induced modulation of recombination or transport is debated [7, 8, 17]. Three-arc spectra have been reported for devices with modified or less efficient contacts, where the additional mid-frequency arc has been linked to a contact or interfacial process whose nature remains unclear [10, 18]. Finally, some devices show an inverted arc at low frequency, also known as negative capacitance or an inductive response. This feature has been linked to both ion-mediated modulation of recombination [10, 19] and charge transport[20, 21]. These debates share a common root: IS alone does not uniquely establish whether a process is associated with recombination or transport, so the same arcs have been interpreted differently.

Intensity-modulated photocurrent (IMPS) and photovoltage (IMVS) spectroscopies offer a complementary route to this question. Instead of modulating the applied voltage and measuring the current, they modulate the intensity of an incident light beam and measure the resulting photocurrent (IMPS) or photovoltage (IMVS). Modulating photogeneration provides new insights into the nature of the underlying processes. IMVS and IMPS have been analyzed mostly independently of one another and of IS, often under different conditions: IMPS typically at short circuit and IMVS at open circuit [22, 23]. IS, IMPS, and IMVS are nevertheless closely related. When measured at the same steady state and within the linear-response regime, their transfer functions are linked so that any one of the three spectra can be calculated from the other two [24].

Equivalent circuits are commonly used to extract quantitative information from impedance spectra. A resistor in parallel with a capacitor reproduces an arc in the Nyquist plot and can therefore represent the corresponding process. However, equivalent circuits do not solve the limitation discussed above. In particular, different equivalent circuits have been proposed for the same IS response of a PSC, implying different physical interpretations, yet they can all reproduce the same IS spectrum equally well [14-16, 25, 26]. Linking the elements of an equivalent circuit to physical processes therefore remains a central challenge.

An equivalent circuit that also reproduces IMVS and IMPS must explicitly include the photogenerated current, and the position of the circuit elements relative to this current can give insights into their nature. The geometry of the equivalent circuit is therefore more tightly restricted. Bertoluzzi and Bisquert proposed theoretically that such an equivalent circuit should describe the same physical mechanisms across IS, IMPS, and IMVS for a photoelectrochemical cell [27]. Ravishankar et al. applied this idea to PSCs, reproducing IS and IMPS spectra with an equivalent circuit containing a photogenerated current, although the parameters obtained from the two techniques showed some discrepancies [28]. Bou et al. later measured IS, IMPS, and IMVS together on PSCs, but fitted only the IS spectra with an equivalent circuit [29]. We proposed a methodology to jointly analyze the three spectra, which we validated on a silicon photodiode [24] and later extended to a $BiVO_4$ photoelectrochemical cell with distributed photogeneration, diffusion, and recombination [30]. This methodology allows a global fit of the three techniques, which fixes the position of each circuit element relative to the photogenerated current. This excludes many equivalent circuit geometries that IS alone cannot distinguish and gives each element a physical meaning.

Here, we show that analyzing the IS, IMPS, and IMVS responses of PSCs together makes it possible to determine whether the processes observed are associated with recombination or transport. We first show conceptually how recombination and transport processes can produce the same IS response, but produce different responses in IMPS and IMVS. We then apply this analysis to three PSCs that show the three common IS responses: a two-arc response, a three-arc response, and an inverted-arc response. This allows us to relate electronic and ion-related processes to their effects on photovoltage and photocurrent, thereby distinguishing recombination-type responses from transport limitations. We also show that it is possible to translate these insights into equivalent circuits that allow for a global fit of the three measurements.

## 2. Results

### 2.1. Mechanisms in Perovskite Solar Cells

**Figure 1**a illustrates the key physical mechanisms occurring in a PSC during operation. An incident photon flux $\phi$, represented with a curly arrow, is absorbed by the perovskite, generating electron-hole pairs. The resulting photogenerated current is given by $j_{ph} = qa\eta_{sep}\phi$, where $q$ is the elementary charge, $a$ is the absorptance and $\eta_{sep}$ is the charge-separation efficiency [24]. The photogenerated carriers are transported through the perovskite and charge-selective contacts, crossing their respective interfaces before extraction, as indicated by the red transport arrow. Not all carriers can be extracted and either recombine within the perovskite bulk or at the perovskite/contact interfaces, as shown by the blue recombination arrows. In addition, ion migration within the perovskite, illustrated in green, is known to occur under illumination or applied bias [31, 32]. The resulting changes in the internal electric field and interfacial carrier densities can, in turn, alter charge-carrier recombination and transport [7, 33-35]. Charge transport, recombination, and ion migration can be coupled to electronic or ionic charge accumulation within the perovskite or at the interfaces[16, 33, 34]. The electron-hole quasi-Fermi-level splitting, shown by the dashed lines, defines the internal voltage and sets an upper limit on the external voltage $V_{ext}$, measured at the device terminals. Transport limitations and non-ideal contacts can reduce $V_{ext}$ below this limit.

IS, IMPS, and IMVS can probe the dynamics of the mechanisms introduced above. Each technique is described by a transfer function relating a small sinusoidal perturbation to the resulting response. Throughout, a tilde denotes the complex amplitude of the corresponding sinusoidal perturbation or response. In IS, a small sinusoidal perturbation of the external voltage $\tilde{V}_{ext}$ is applied and the resulting external current response $\tilde{j}_{ext}$ is measured. The IS transfer function is defined as $Z = \tilde{V}_{ext}/\tilde{j}_{ext}$. In IMPS and IMVS, the perturbation is instead applied to the photon flux $\tilde{\phi}$ and the resulting $\tilde{j}_{ext}$ and $\tilde{V}_{ext}$ responses are measured, respectively. Using the convention in which $\tilde{j}_{ext}$ is positive for a positive $\tilde{V}_{ext}$, these transfer functions are defined as $Q = -\tilde{j}_{ext}/q\tilde{\phi}$ for IMPS and $W = \tilde{V}_{ext}/q\tilde{\phi}$ for IMVS. The three transfer functions are not independent but are related through

$$Z = \frac{W}{Q} \tag{1}$$

The explicit frequency dependence is omitted for clarity, but $Z$, $Q$, and $W$ are all complex functions of frequency, and Equation (1) applies at each frequency. This relation applies when the three measurements probe the same steady state and the perturbations are sufficiently small to ensure a linear relationship between input and output. Equation (1) therefore provides an experimental consistency test, since any one of the three spectra can be reconstructed from the other two [24].

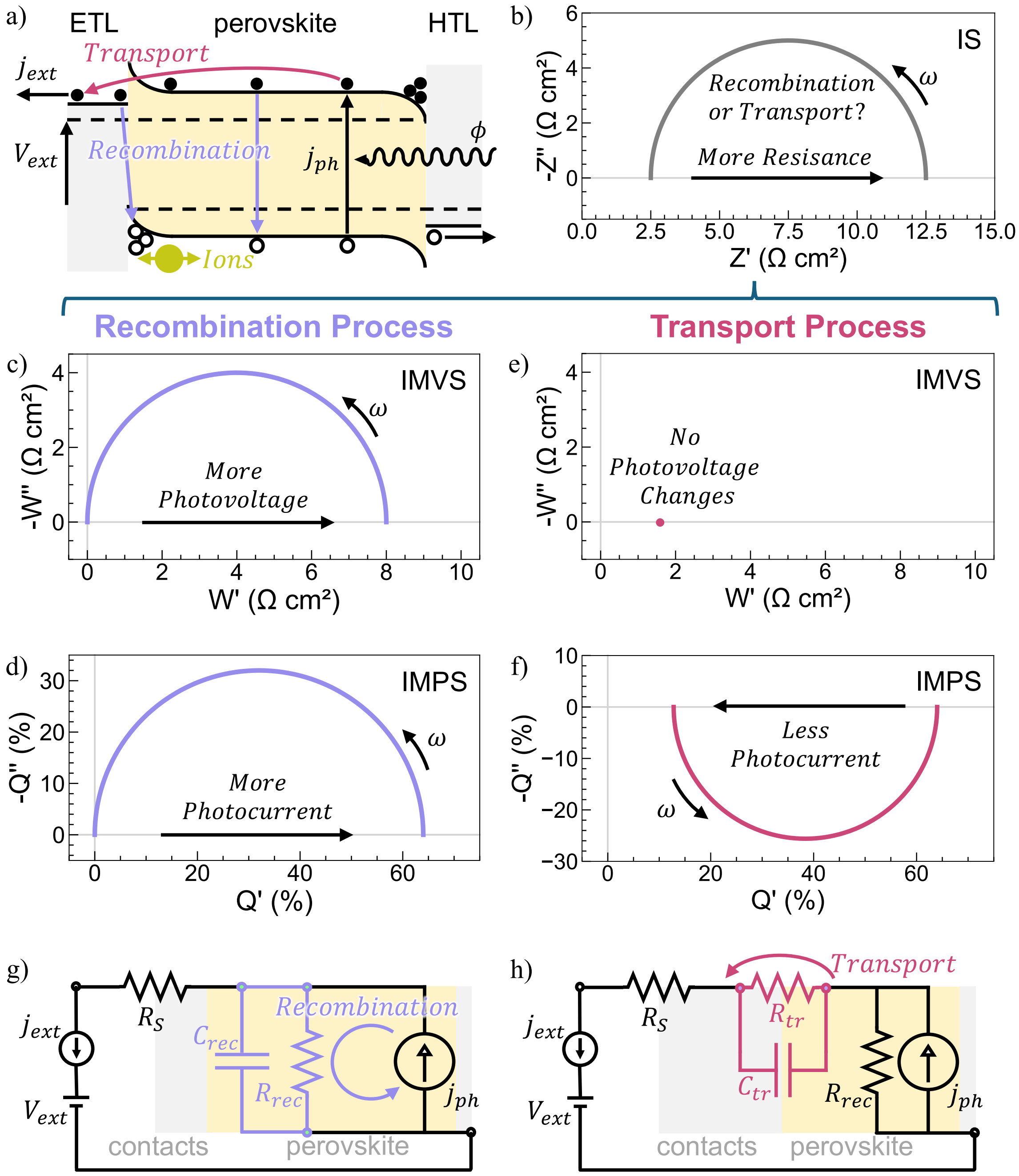

**Figure 1**. a) Key physical mechanisms in a PSC during operation: photogeneration ($j_{ph}$) driven by the photon flux ($\phi$), transport (red), recombination (blue), and ion migration (green). The dashed lines are the electron and hole quasi-Fermi levels, whose splitting defines the internal voltage. $j_{ext}$ and $V_{ext}$ are the external current and voltage. b) Idealized IS Nyquist plot of a process in which charge accumulation is coupled to either recombination or transport. c) IMVS and d) IMPS spectra for the recombination case. e) IMVS and f) IMPS spectra for the transport case. Simplest equivalent circuits for the g) recombination ($R_{rec}$, blue, in parallel with $j_{ph}$) and h) transport ($R_{tr}$, red, in series with $j_{ph}$) cases, each with a capacitor in parallel

with the resistor. Both reproduce the IS spectrum in b), but g) also reproduces c) and d), while h) reproduces e) and f).

**Figure 1**b illustrates the idealized IS response of a process, either recombination or transport, observed as an arc whose characteristic frequency $f_c$, at the top of the arc, falls within the range of measured frequencies. This frequency is related to the characteristic time by $\tau = (2\pi f_c)^{-1}$. At high frequencies (short times), the modulated charge cannot complete the process within a modulation cycle, so the voltage reverses before appreciable modulated charge can accumulate. Then the impedance approaches a purely real high-frequency limit that includes the series resistance $R_S$, which could account for the fast ohmic contributions from the electrodes, contacts, transport layers, and electrical connections. As the frequency decreases, more modulated charge accumulates during each cycle and a progressively larger fraction of the current flows through the recombination or transport pathway. The charging and discharging delay the response, producing a phase shift between voltage and current, while the resistance of the recombination or transport pathway progressively contributes to the real part of the impedance, generating an arc in the Nyquist plot. At sufficiently low frequencies (long times), accumulation follows the perturbation quasi-instantaneously, the charging current becomes negligible, and the impedance approaches a purely real limit equal to the high-frequency $R_S$ plus the resistance of the recombination or transport pathway.

IS can thus separate processes with sufficiently different characteristic times and extract an effective resistance and capacitance for each. However, because both the perturbation and response are purely electrical, with no modulation of photogeneration, an arc alone cannot reveal whether the underlying process is associated with recombination or transport. In both cases, the charges first accumulate and then cross a resistive pathway, and IS measures only the corresponding changes in current at the terminals, so both cases can generate the same IS response shown in **Figure 1**b.

Although IS cannot distinguish between recombination and transport, IMVS and IMPS can, because modulating the photogeneration changes how each process contributes to the response. **Figure 1**c and **Figure 1**d show the IMVS and IMPS spectra, respectively, of the process in which the modulated carrier density leads to a modulation in the recombination, which generates the IS arc in **Figure 1**b. The IMVS spectrum (**Figure 1**c) closely resembles the IS spectrum in shape, showing an arc, but without the $R_S$ offset. Since no AC current is extracted during an IMVS measurement, series elements produce no AC voltage drop, and $W$ is independent of any series impedance [24]. The IMPS spectrum (**Figure 1**d) also exhibits an arc.

This process of recombination coupled with accumulation thus generates an arc in all three techniques. It creates an arc in IMVS and IMPS because recombination acts directly on the photogenerated carriers, so the coupled accumulation is that of those carriers. This accumulation can be electronic, associated with the change in carrier density with the quasi-Fermi-level splitting (chemical capacitance), or dielectric, associated with material polarization between the contacts, including the absorber (geometric capacitance)[16]. For IMVS, at sufficiently high frequencies, the modulated photogenerated charge cannot accumulate appreciably before the modulated photon flux reverses. The modulated internal voltage therefore tends to zero, and so does $W$. As the frequency decreases, more modulated photogenerated charge accumulates during each cycle, and a modulated photovoltage builds up, increasing $W$. As in IS, charging and discharging the corresponding capacitance delays the photovoltage relative to the photon flux, producing a phase shift in $W$ and generating the IMVS arc. At sufficiently low frequencies, accumulation follows the modulation quasi-statically, and $W$ becomes real, closing the arc. In IMPS, the internal photovoltage drives photogenerated charge extraction through $R_S$, competing with recombination. Following the same reasoning, at sufficiently high frequencies, there is no appreciable modulated internal voltage to drive extraction, so the AC current and $Q$ also tend to zero. As frequency decreases, the extracted current rises as the internal photovoltage builds up, also generating an arc that closes when $Q$ approaches the steady-state extracted fraction of photogenerated charge (**Figure 1**d).

**Figure 1**e and **Figure 1**f show the IMVS and IMPS spectra, respectively, of the process in which the modulated carrier density leads to a modulation in the transport, which generates the same IS response in **Figure 1**b. Here it is important to note that recombination is always present: part of the photogenerated charge recombines, and the remaining charges encounter the transport process on their way to extraction. In IMVS, by definition, no modulated charge is extracted, so all of it eventually recombines without crossing the transport pathway. The IMVS response (**Figure 1**e) is therefore set only by photogeneration and recombination, and is unaffected by the transport resistance, its associated accumulation, and $R_S$. In this case, $W$ is real and independent of frequency, so no arc appears. In IMPS (**Figure 1**f), the extracted current again results from the competition between extraction and recombination, but now the extraction path includes the transport process. At sufficiently high frequencies, the modulated charge accumulates on either side of the transport resistance without appreciable opposition,

because the voltage built up by this accumulation remains small. The modulated current is then carried by the charging and discharging of this accumulation rather than by charge flowing through the resistance. From the external terminals, it is as if the charge were extracted through $R_S$ alone. For example, if the transport limitation is the low conductivity of a selective contact, the modulated charge accumulates at both of its interfaces without major opposition at high frequency, yielding an effective extracted photocurrent. As the frequency decreases, the accumulated charge progressively opposes further accumulation, leading to a phase shift. The extracted charge also has to flow through the transport resistance in addition to $R_S$. The extraction path thus changes from $R_S$ at high frequencies to $R_S$ plus the transport resistance at low frequencies. Because recombination is unchanged, a larger fraction of the photogenerated charge recombines at lower frequencies and $Q$ decreases, producing an inverted arc (**Figure 1**f).

Equivalent circuits provide a compact way to represent and fit different responses, including these two limiting cases. In our previous work, we introduced a general equivalent circuit, whose general impedances can represent different mechanisms, including the mechanisms in **Figure 1**a, together with the corresponding expressions for $Z$, $W$ and $Q$ [24]. Here we use its simplest form, in which a process that generates an arc in IS as in **Figure 1**b is represented by a resistor in parallel with a capacitor. We denote the resistor $R_{rec}$ when the process is recombination and $R_{tr}$ when it is transport. What distinguishes the two resistances is the position of this element relative to the photogenerated current source $j_{ph}$. This position matters only in IMVS and IMPS, where $j_{ph}$ is modulated. In IS, photogeneration is not modulated, so $j_{ph}$ has no AC component and the equivalent circuits used to analyze IS usually omit it. If the resistance is placed in parallel with $j_{ph}$ (**Figure 1**g), it provides a path through which photogenerated charge recombines inside the device without being extracted. If it is placed in series with $j_{ph}$ and the terminals (**Figure 1**h), the extracted charge must flow through it, and it represents the resistance accounting for transport limitations. Here, transport can include interfacial charge transfer and transport within selective layers, with the associated capacitance encompassing dielectric and interfacial contributions. There is always some recombination, so the transport circuit also includes $R_{rec}$ in parallel with $j_{ph}$ (**Figure 1**h). Both equivalent circuits reproduce the IS arc in **Figure 1**b, so IS cannot distinguish between them, but IMVS and IMPS can, because of the physical differences described above. Further discussion of the equivalent circuits is given in the Supporting Information.

**Figure 1** presents two limiting single-process cases, not every possible PSC response. Throughout, a recombination-type response includes charge storage coupled to recombination and does not necessarily imply changes in the underlying recombination characteristics, such as trap densities or recombination coefficients. A transport-type response refers here to a limitation confined to the series extraction path. Real devices contain several electronic and ion-coupled processes, and the ionic redistribution may modify either recombination or transport.

In the following sections, we apply this analysis to three PSCs with typical IS spectra, showing two arcs, three arcs, and an inverted arc, to identify the processes behind each arc, such as recombination, transport, and the effect of the ion-related processes on them.

### 2.2. High-Efficiency PSC: Two-Arc Impedance

We measure IS, IMPS, and IMVS on a high-efficiency PSC to resolve the processes behind its two-arc impedance response, specifically whether each is associated with recombination or transport. **Figure 2**a shows a schematic of the PSC, based on a triple-cation $Cs_{0.05}(MA_{0.05}FA_{0.95})_{0.95}Pb(Br_{0.05}I_{0.95})_3$ perovskite passivated with $EDAI_2$–PEAI in an inverted device architecture, together with its current-voltage (JV) curve. The device exhibits a power conversion efficiency (PCE) of 24.3 % with an open-circuit voltage ($V_{OC}$) = 1.200 V, short-circuit current density ($J_{SC}$) = 23.5 mA cm$^{-2}$, and fill factor (FF) = 80.8 % [36]. Details of the JV measurement conditions are provided in the **Experimental Section**. A fundamental requirement for combining IS, IMPS, and IMVS is that all three techniques are measured under the same steady-state conditions, here at open circuit under the same DC illumination, as indicated by a cross on the JV curve in **Figure 2**a. A blue LED with a narrow, well-defined emission spectrum was used as the illumination source (see **Experimental Section**). The measured IS, IMVS, and IMPS spectra are shown as black dots in **Figure 2**b, **Figure 2**c, and **Figure 2**d, respectively. Because the three transfer functions are related by Equation (1), any one of them can be calculated from the other two. The gray circles in **Figure 2**c show the IMVS spectrum calculated from the IS and IMPS spectra using that equation. Their close agreement with the measured IMVS spectrum indicates that Equation (1) holds, so we can rely on the data for the analysis that follows.

The IS spectrum (**Figure 2**b) shows two arcs, the most commonly reported response for high-efficiency PSCs [14-16]. Their characteristic frequencies, taken at the maximum of each arc, are about 100 kHz and 1 Hz, corresponding to characteristic times of about 1.6 µs and 160 ms. Such widely separated time scales are usually attributed to an electronic process at high frequencies and an ion-related process at low frequencies, because ion migration is orders of magnitude slower than electronic responses [37, 38]. This assignment from IS rests on the characteristic times and the literature, but does not establish whether either arc is associated with recombination or transport, as discussed in Section 2.1. We therefore turn to IMVS and IMPS measured under the same steady state.

The IMVS spectrum (**Figure 2**c) shows two arcs with the same characteristic frequencies as the IS arcs, suggesting that they originate from the same two processes. As discussed in Section 2.1 (**Figure 1**c and **Figure 1**e), since no AC current is extracted in IMVS, only processes that act on the internal voltage can contribute to it, so both processes behave as recombination-type responses. The IMPS spectrum (**Figure 2**d) also shows two arcs, none of them inverted, unlike the arc in **Figure 1**f, which excludes a transport process. The IMPS arcs appear at higher characteristic frequencies (about 250 kHz and 2.5 Hz) and with different relative sizes than those of IS and IMVS. This is expected, since in IMPS the modulated charges are extracted through series resistance, competing with recombination, so the same processes are probed differently. The global fit described below reproduces these frequency shifts using the same circuit parameters for all three spectra, supporting their assignment to the same processes. Both processes therefore have the same functional role: they are associated with recombination, and no transport contribution is resolved beyond that included in the series resistance.

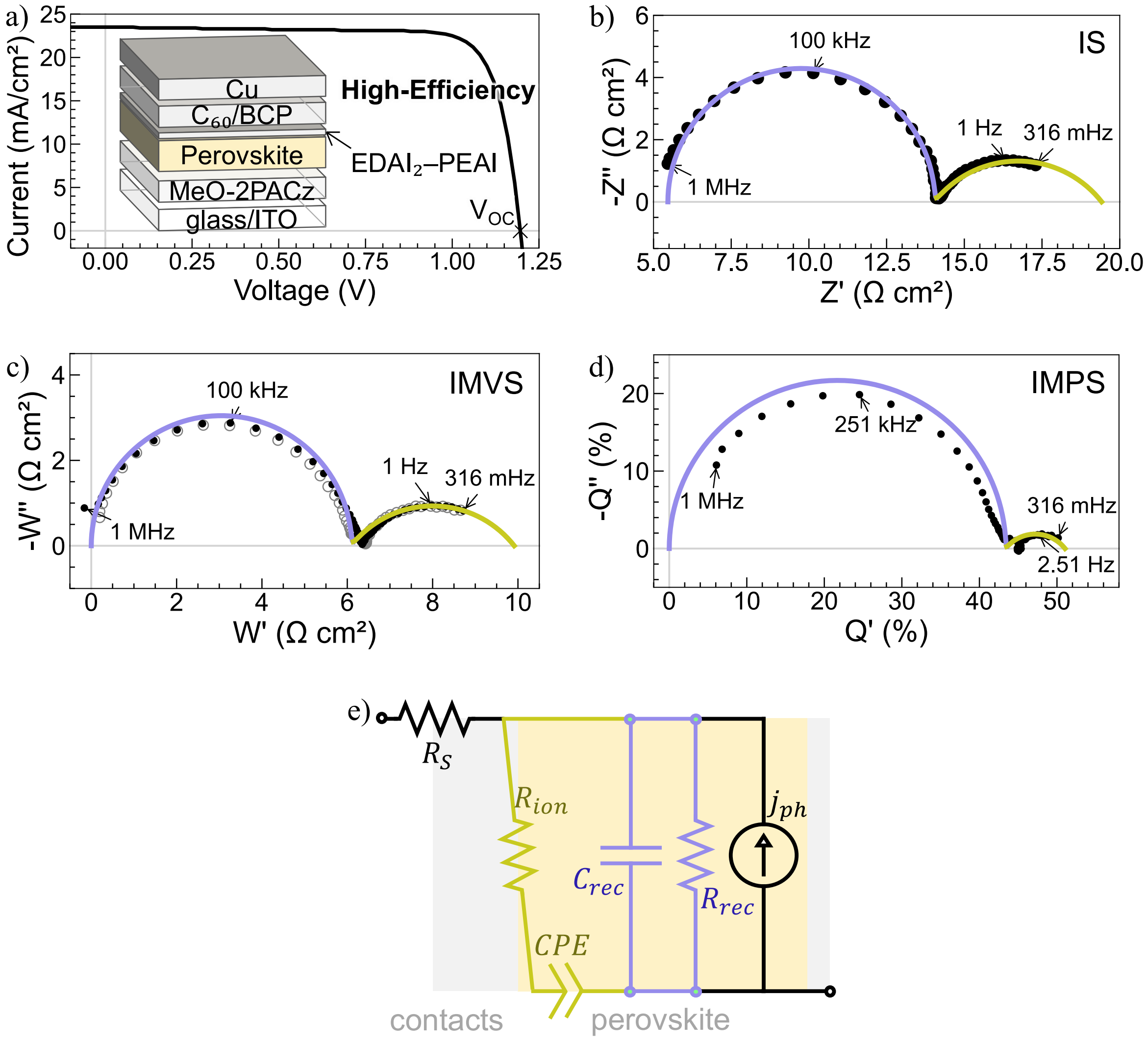

**Figure 2**. a) Measured JV curve of the investigated PSC, with a schematic of the device structure in the inset. The cross marks the open-circuit condition at which all spectra were measured. b) IS, c) IMVS, and d) IMPS spectra (black dots). The gray circles in c) are the IMVS spectrum calculated from the IS and IMPS spectra using Equation (1). Lines in b) to d) show the global fit of the three spectra with the equivalent circuit in e), colored by the process that generates each arc: blue for the high-frequency arc and green for the low-frequency arc. Markers indicate the characteristic frequency of each arc and the maximum and minimum measured frequencies. e) equivalent circuit used to fit the data.

A key observation is the effect of the slow, ion-related process. In IS, it increases the resistance toward low frequencies, which, as discussed in Section 2.1, could reflect either a transport-type or a recombination-type response. IMVS and IMPS show that both the photovoltage and the extracted photocurrent increase toward low frequencies, the signatures of a recombination-type response as discussed in Section 2.1. One possible origin is ionic charge accumulation coupled to recombination, producing a response analogous to that associated with geometric or chemical capacitance. The same trends are also consistent with ionic redistribution suppressing recombination. Both contributions may coexist. This is important because ions critically influence perovskite solar cell performance, yet the underlying mechanisms remain under debate [7-10]. Combining IMVS and IMPS with IS helps to clarify the role of this process through a direct observation of the corresponding photovoltage and photocurrent changes. Since the low-frequency limit corresponds to the steady state, the interpretation of the low-frequency ion-related response suppressing recombination is consistent with the high $V_{OC}$ of this device.

To represent and quantify these responses, we use the equivalent circuit in **Figure 2**e, which places $R_{rec}$ and a branch consisting of $R_{ion}$ in series with a constant phase element ($CPE$) in parallel with $j_{ph}$. Since no transport process is resolved, this equivalent circuit treats all transport contributions through the series resistance $R_S$.

The $R_{ion}$-$CPE$ branch can represent ionic charge accumulation coupled to recombination. As frequency decreases, the AC charging current through this branch diminishes, allowing larger photovoltage and extracted photocurrent responses. Notably, the same branch can also describe ion-mediated suppression of recombination. In this interpretation, an additional charge recombination path is present at frequencies above the ionic relaxation and is progressively suppressed as ionic redistribution follows the perturbation toward lower frequencies, because the $CPE$ in the ionic branch behaves as a short circuit and an open circuit in the high- and low-frequency limits, respectively. Further discussion of the equivalent circuits is given in the Supporting Information.

We fitted the three spectra globally with this equivalent circuit, as shown by the lines in **Figure 2**b–d. The fit reproduces all three spectra well, supporting the physical picture outlined above. Because IMVS and IMPS modulate photogeneration, the combined analysis fixes the position of $j_{ph}$ in the circuit, which IS alone cannot do [24], and this position is what distinguishes recombination from transport. The global fit also gives access to the product $\eta_{sep} \cdot a$, and from it the differential EQE and IQE and the charge-collection efficiency $\eta_{col}$ (see **Table S1**). We now apply the same analysis to a device with non-optimized contacts.

### 2.3. Non-Optimized Contacts PSC: Three-Arc Impedance

Next, we illustrate the power of this methodology to identify the loss processes in less efficient devices. **Figure 3**a shows the schematic of a second PSC, otherwise identical to the device in **Figure 2** but with non-optimized contacts: no $EDAI_2$–PEAI passivation interlayer and a thinner copper contact (~20 nm) [36]. The JV curve is also shown in **Figure 3**a; the device PCE = 18.81 % is lower than that of the device in **Figure 2**, with $V_{OC}$ = 1.118 V, $J_{SC}$ = 24.10 mA $cm^{-2}$, and FF = 72.82 %. As before, all three techniques were measured at open circuit under the same DC illumination (cross in **Figure 3**a). The measured IS, IMVS, and IMPS spectra are shown as black dots in **Figure 3**b, **Figure 3**c, and **Figure 3**d, respectively. The gray circles in **Figure 3**d show the IMPS spectrum calculated from the IS and IMVS spectra using Equation (1). Also for this device, they agree well with the measured spectrum over the whole frequency range.

The IS spectrum (**Figure 3**b) shows three arcs, a response previously reported for PSCs with modified or less efficient selective contacts and associated with an additional intermediate-frequency process [18, 39]. However, whether the associated resistance reflects a transport limitation or a change in recombination was not conclusively established. The characteristic frequencies, taken at the maximum of each arc, are about 200 kHz, 13 kHz, and 5 Hz, corresponding to characteristic times of about 0.8 μs, 12 μs, and 32 ms. The high- and low-frequency arcs lie at time scales comparable to those of the device in **Figure 2** (about 100 kHz and 1 Hz) and are therefore also attributed to an electronic and an ion-related process at high and low frequencies, respectively. The additional arc appears at an intermediate frequency far above the time scales usually associated with ion migration, so it likely corresponds to an electronic response. As discussed in Section 2.1, We use IMVS and IMPS measured under the same steady state to establish the nature of this process.

The IMVS spectrum (**Figure 3**c) shows only two arcs, at about 200 kHz and 4 Hz, matching the high- and low-frequency IS arcs, whereas the IMPS spectrum (**Figure 3**d) shows three, at about 300 kHz, 20 kHz, and 5 Hz, close to the three IS arcs. The mid-frequency arc in IMPS is inverted and has no counterpart in IMVS, which are the signatures of a transport process (see also **Figure 1**). The high- and low-frequency processes appear as non-inverted arcs in both techniques, so, as in the device in **Figure 2**, they behave as recombination-type responses. As above, the low-frequency process increases both the photovoltage and the photocurrent, consistent with ionic charge accumulation and/or reduced recombination associated with ionic redistribution. The additional intermediate-frequency process in IS is therefore a transport limitation. This process appears in the device with non-optimized contacts, suggesting that contact optimization may avoid it, as in the device in **Figure 2**.

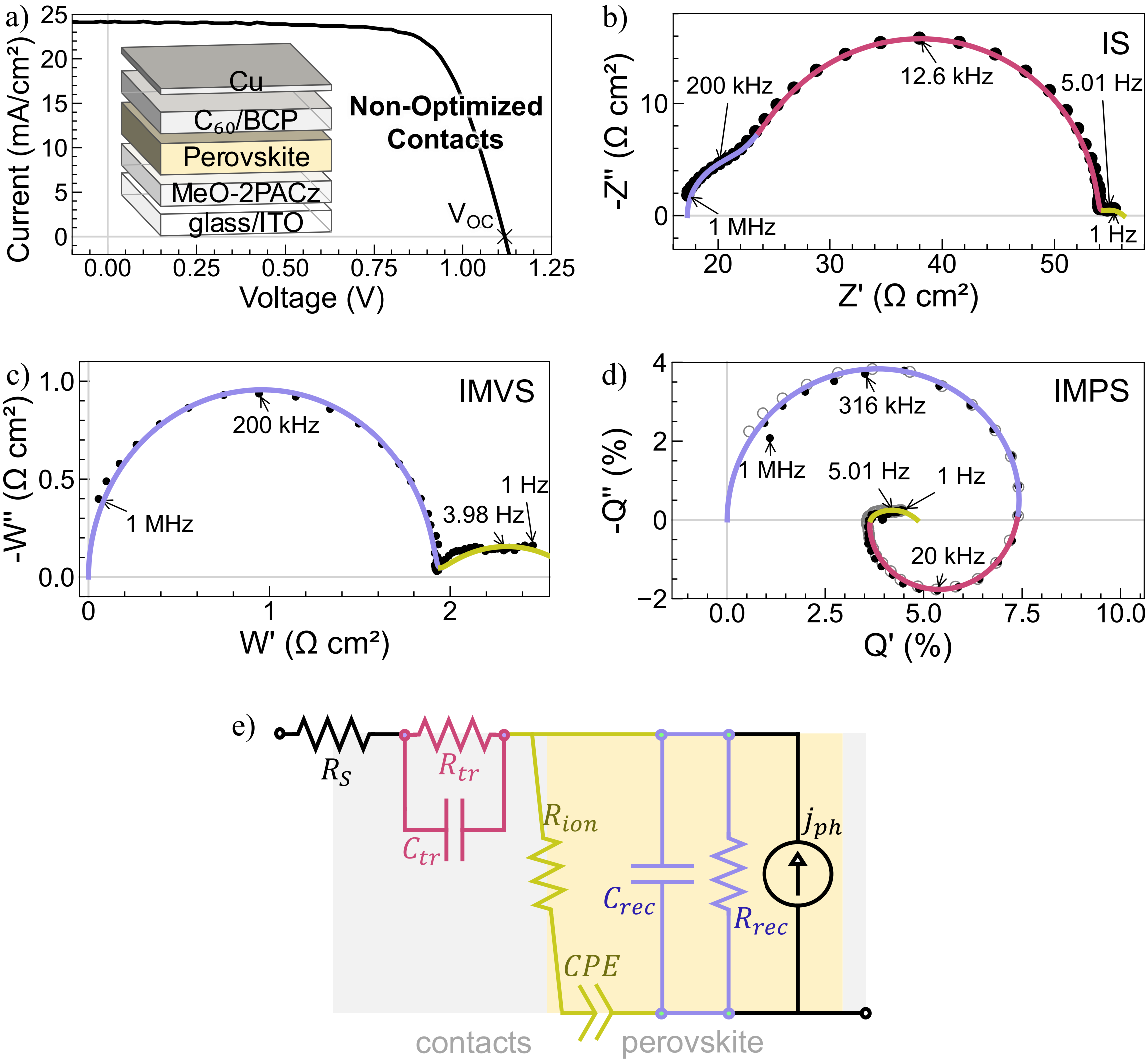

**Figure 3**. a) Measured JV curve of the investigated PSC with non-optimized contacts, device structure in the inset. The cross marks the open-circuit condition at which all spectra were measured. b) IS, c) IMVS, and d) IMPS spectra (black dots). The gray circles in d) are the IMPS spectrum calculated from the IS and IMVS spectra using Equation (1). Lines in b) to d) show the global fit of the three spectra with the equivalent circuit in e), colored by the process that generates each arc: blue for the high-frequency arc, red for the mid-frequency arc, and green for the low-frequency arc. Markers indicate the characteristic frequency of each arc and the maximum and minimum measured frequencies. e) Equivalent circuit used to fit the data.

We represent these responses with the equivalent circuit in **Figure 3**e, which is the equivalent circuit in **Figure 2**e with an additional RC pair, $R_{tr}$ in parallel with a capacitor $C_{tr}$ (red), placed in series with $j_{ph}$ and the terminals. This combines the two cases of **Figure 1**g and **Figure 1**h, with transport in series and recombination in parallel with $j_{ph}$. Because $R_{tr}$ lies in the extraction path, it lowers the extracted photocurrent but does not affect the photovoltage. We fitted the three spectra globally with this equivalent circuit (lines in **Figure 3**b–d, with each arc colored as its element), which closely reproduces all key features of the experimental data, including the inverted IMPS arc. The extracted parameters are summarized in **Table S1**. The additional transport limitation, which may arise from charge transport through a selective layer or charge transfer at an interface, is consistent with the lower FF of this device (72.8%, compared with 80.8% for the device in **Figure 2**). The lower $R_{rec}$ and $R_{ion}$ are consistent with increased interfacial recombination associated with the absence of the passivation interlayer [36], resulting in a lower $V_{OC}$. Further discussion of the equivalent circuit is given in the Supporting Information.

### 2.4. Lower-Efficiency PSC: Inverted-Arc Impedance

Finally, we consider a third PSC whose IS spectrum shows an inverted low-frequency arc. **Figure 4**a shows the schematic of this device, which differs from those in **Figure 2** and **Figure 3** in the perovskite composition ($Cs_{0.05}FA_{0.87}MA_{0.08}PbI_3$ instead of $Cs_{0.05}(MA_{0.05}FA_{0.95})_{0.95}Pb(Br_{0.05}I_{0.95})_3$), the electron-transport layer (PCBM instead of $C_{60}$), and the metal contact (silver instead of copper), and has no passivation interlayer. The device was characterized after shipping and storage, with the JV curve shown in **Figure 4**a, corresponding to a PCE = 16.77 %, with $V_{OC}$ = 1.070 V, $J_{SC}$ = 20.86 mA cm$^{-2}$, and FF = 75.14 %, all lower than for the device in **Figure 2**. As before, all three techniques were measured at open circuit, under the same DC illumination (cross in **Figure 4**a). The measured IS, IMVS, and IMPS spectra are shown as black dots in **Figure 4**b, **Figure 4**c, and **Figure 4**d, respectively. The IMVS spectrum calculated from the IS and IMPS spectra using Equation (1) agrees well with the measured spectrum at high frequencies, and less well at low frequencies, where the small, noisy low-frequency feature in IMPS limits the reconstruction.

The IS spectrum (**Figure 4**b) shows an arc at high frequencies followed by an inverted arc at low frequencies, also known as negative capacitance or inductive response, which has been reported extensively in PSCs and other photovoltaic technologies [40-43]. The characteristic frequencies, taken at the extremum of the imaginary component of each arc, are about 250 kHz and 250 Hz, corresponding to characteristic times of about 0.6 μs and 0.6 ms. The high-frequency arc lies at a time scale similar to those of the devices in **Figure 2** and **Figure 3**, whereas the low-frequency process is considerably faster than their ion-related processes. This shorter time scale does not exclude ionic involvement, since millisecond ionic responses have also been reported [10, 37] and the different perovskite composition merits considering a different ionic mobility. In fact, often the low-frequency inverted arc is assigned to an additional recombination pathway activated by ion migration [6, 19, 44]. Whether ionic or not, in line with the discussion in Section 2.1, IS alone cannot establish whether a resistance that decreases toward low frequencies reflects an increase in recombination or a decrease in a transport limitation. We therefore turn to IMVS and IMPS measured under the same steady state.

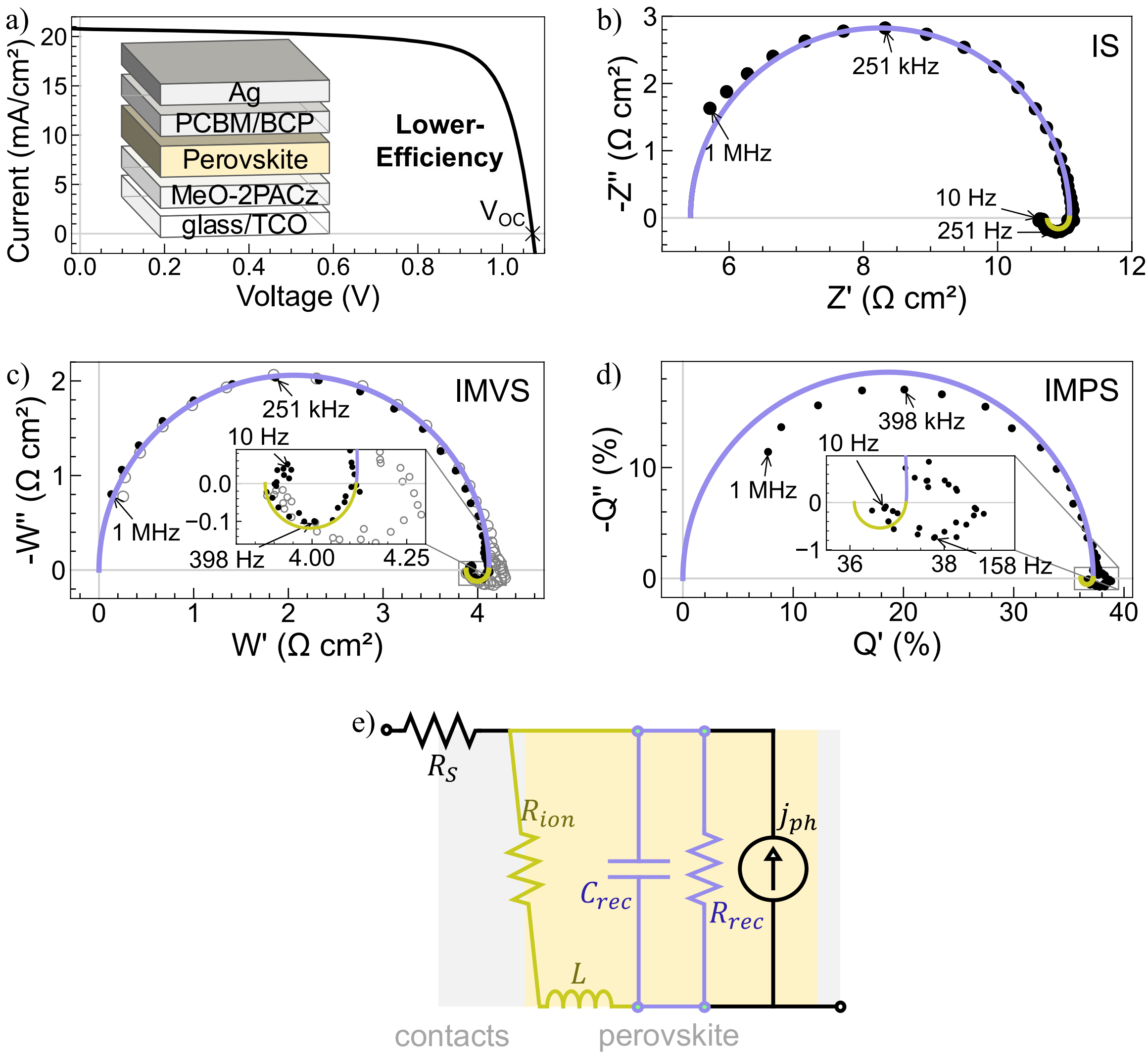

**Figure 4**. a) Measured JV curve of the lower-efficiency PSC, showing lower performance than the devices in **Figure 2** and **Figure 3**, with a schematic of the device structure in the inset. The cross marks the open-circuit condition at which all spectra were measured. b) IS, c) IMVS, and d) IMPS spectra (black dots). The gray circles in c) are the IMVS spectrum calculated from the IS and IMPS spectra using Equation (1). Lines in b) to d) show the global fit of the three spectra with the equivalent circuit in e), colored by the process that generates each arc: blue for the high-frequency arc and green for the low-frequency (inverted) arc. Markers indicate the characteristic frequency of each arc and the maximum and minimum measured frequencies. e) Equivalent circuit used to fit the data.

The IMVS spectrum (**Figure 4**c) shows the same two arcs as IS, at about 250 kHz and 400 Hz, with the low-frequency arc also inverted, so the photovoltage decreases toward low frequencies. As discussed in Section 2.1, a transport process would leave IMVS unchanged, so the presence of this arc excludes it and indicates that the low-frequency process acts on recombination. The IMPS spectrum (**Figure 4**d) shows a clear high-frequency arc at about 400 kHz and a small low-frequency feature around 150 Hz, consisting of a small inverted arc overlapping with a shift toward more positive $Q'$ values. The inverted arc is consistent with a reduced photocurrent but is too small to be resolved from the noise, and we attribute the shift to a minor device instability or a measurement artifact. Therefore, while IS shows a lower resistance at low frequencies, which could reflect more recombination or less transport limitation, IMVS shows a lower photovoltage, and IMPS is consistent with a lower photocurrent once the low-frequency process is activated, indicating that this process in fact increases recombination. The associated decrease in effective recombination resistance is consistent with increased losses and the lower $V_{OC}$ of this device compared with the devices in **Figure 2** and **Figure 3**. Notably, in those devices, the low-frequency ion-related responses were consistent with *ionic accumulation* and/or *reduced recombination* once the process was activated, whereas here it *increases recombination*.

We represent these responses with the equivalent circuit in **Figure 4**e, which is the equivalent circuit in **Figure 2**e with an inductor $L$ replacing the $CPE$ in the low-frequency branch ($R_{ion}$ in series with $L$, green). Once the low-frequency process is activated, this branch opens an additional recombination channel in parallel with $R_{rec}$, increasing the total recombination [19]. This produces the inverted arc in IS and IMVS, and it also reduces the extracted photocurrent in IMPS, although this effect is small in the measured spectrum. We fitted the three spectra globally with this equivalent circuit (lines in **Figure 4**b–d, with each arc colored as its element), which reproduces the spectra closely across the full frequency range, including the inverted arc in IS and IMVS. The only exception is the small low-frequency IMPS feature: the fit predicts a small inverted arc, consistent with the low-frequency limit, but not the shift toward more positive $Q'$, in line with the artifact interpretation above. The extracted parameters are summarized in **Table S1**. Further discussion of the equivalent circuit is given in the Supporting Information.

## 3. Discussion

In the three devices, the combination of IS with IMVS and IMPS associates each feature observed in IS with recombination or transport. In all three devices, the high-frequency arc behaves as a recombination-type response. The low-frequency responses are consistent with ionic charge accumulation and/or reduced recombination in the two-arc and three-arc devices, where both photovoltage and photocurrent increase, whereas the decrease in photovoltage and the photocurrent response in the inverted-arc device support increased recombination. The three-arc device additionally shows a transport process, which appears in IS as one more arc, indistinguishable by shape from the recombination-related arcs, but is identified as transport because it is absent in IMVS and produces an inverted arc in IMPS. Across the devices, the IMVS arcs coincide with the IS arcs in characteristic frequency, whereas the IMPS arcs are displaced by modest factors, as expected from the competition between extraction through the series resistance and recombination in IMPS.

The methodology establishes the functional role of each process, recombination or transport, but not its microscopic nature, including where it occurs, whether the associated capacitance is chemical or geometric, and whether it is electronic or ionic. The ion-related assignment of the low-frequency processes, in particular, rests on characteristic times and previous work. To obtain further information, particularly to address these questions, the methodology could be extended to IS, IMPS, and IMVS measured at different voltages, light intensities, and temperatures, and combined with other techniques, with IMPLS being particularly promising [45-47].

The limitation of IS that motivates this work is inherent to the technique, not to the model used to interpret it. Interpreting IS data using equivalent circuits, drift-diffusion simulations, or analytical models all face the same limitation, because IS probes the electrical response without perturbing photogeneration, so different sets of parameters or assumptions can reproduce the same spectrum. Adding IMPS and IMVS data clarifies this ambiguity independently of the model.

## 4. Conclusions

IS can resolve dynamic processes by their characteristic times, but it cannot establish whether a measured process is associated with recombination or transport. To address this, we combined IS with IMVS and IMPS, which modulate photogeneration and show directly how each process changes the photovoltage and photocurrent. These changes allow the resistance changes observed in IS to be associated with recombination or transport. We applied this methodology to three PSCs with commonly-seen, distinct IS responses: a two-arc response, a three-arc response, and an inverted-arc response. For each PSC, we globally fitted the three spectra with an equivalent circuit that places the photogenerated current explicitly relative to the elements representing each process, defining their nature as transport or recombination.

In the two-arc device, both processes were associated with recombination, the high-frequency arc with an electronic process and the low-frequency arc with an ion-related process consistent with ionic charge accumulation and/or reduced recombination, as shown by the increase in photovoltage and photocurrent. In the three-arc device, the additional mid-frequency process was instead identified as a transport limitation; strikingly, this same process appears as another arc in IS, is absent in IMVS, and appears as an inverted arc in IMPS, illustrating how a single process can manifest entirely differently across the three techniques, and how IS alone can therefore be insufficient to assign it correctly. In the inverted-arc device, the low-frequency process was also associated with ionic migration in the perovskite, but here it generated an additional recombination pathway, increasing total recombination and reducing resistance, photovoltage, and photocurrent.

In this work, we have shown that combining different modulation techniques can unveil the nature of processes in PSCs. By addressing the long-standing limitation of uniquely assigning processes using IS, our work provides a clearer understanding of the mechanisms governing PSC operation, enabling the processes that affect performance, whether limiting or beneficial, to be identified with greater confidence, and ultimately supporting efforts to overcome the major remaining challenges of this technology.

## Experimental Section

*Device Fabrication*

The PSCs shown in **Figure 2** and **Figure 3**, whose respective device architectures are presented in the corresponding figures, had an active area of 0.06 $cm^2$ and 0.16 $cm^2$, respectively, and were fabricated at the University of Potsdam following the procedure reported by Seid et al. [36].

The PSC shown in **Figure 4**, whose respective device architecture is presented in the corresponding figure, had an active area of 0.048 $cm^2$ and was fabricated at the University of Pavia following the procedure reported by Degani et al. [36, 48, 49].

*Current Density–Voltage Curve*

The current density–voltage (JV) curves of the PSCs presented in **Figure 2** and **Figure 3** were measured at the University of Potsdam at 25 °C in a nitrogen-filled glovebox using a Keithley 2400 source meter. The measurements were performed under standard AM 1.5G illumination, provided by a Class AAA solar simulator equipped with an Oriel xenon lamp. The illumination intensities were 93.8 and 104.3 mW $cm^{-2}$ for the devices in **Figure 2** and **Figure 3**, respectively. Both forward and reverse scans were recorded over a voltage range of -0.2 to 1.2 V at a scan rate of 20 mV $s^{-1}$.

The JV curve of the PSC presented in **Figure 4** was measured at AMOLF from -0.05 to 1.2 V using a G2V Pico solar simulator and a Keithley source-measure unit. The measurements were performed inside a glovebox at room temperature without an aperture mask. The illumination intensity was calibrated to 1 sun using a calibrated reference photodiode. In addition, the output of each LED was measured individually using a Thorlabs PM100D optical power meter.

*Frequency-Domain Characterization*

IS, IMPS, and IMVS measurements were performed using a PAIOS system (Fluxim AG). Illumination during the small-perturbation measurements was provided by a 465 nm blue LED (XP-E2, Cree LED). Frequency-dependent changes in the amplitude and phase of the LED electroluminescence modulation were corrected following the previously reported protocol [50]. The measurements were performed inside a glovebox at room temperature without an aperture mask. The measurements corresponding to **Figure 2**, **Figure 3**, and **Figure 4** were performed under DC steady-state conditions of open-circuit voltage and illumination intensities of 14.18, 37.83, and 37.22 mW $cm^{-2}$, respectively.

The AC amplitudes were selected for each measurement to maintain a linear response while providing an adequate signal-to-noise ratio. For IS, voltage-modulation amplitudes of 20 mV were used for **Figure 2** and **Figure 3**, while 10 mV was used for **Figure 4**. For IMPS, the light-modulation amplitude was 20% of the DC illumination intensity for **Figure 2** and **Figure 3** and 7% for **Figure 4**. For IMVS, the light-modulation amplitude was 15% of the DC illumination intensity for **Figure 2**, 20% for **Figure 3**, and 7% for **Figure 4**.

**Supporting Information**

Details of the equivalent-circuit models and analysis, together with the parameters obtained from fitting.

**Conflicts of interest**

The authors declare no conflicts of interest.

**Acknowledgements**

The work of A.O.A. is part of the project Faculty of Impact (FoI.2025.022), grant ID DUEYX74140, financed by the Dutch Research Council (NWO) and carried out at the research institute AMOLF. A.O.A. and B.E. acknowledge funding from the European Research Council (ERC) under the European Union's Horizon 2020 research and innovation program under grant agreement No. 947221. A.S. acknowledges funding from SolarLab within SolarNL, a national research, innovation and industrial development program funded by the Netherlands National Growth Fund. The work is part of the Dutch Research Council NWO and was performed at the research institute AMOLF. F.L. and B.A.S. acknowledge funding from the Volkswagen Foundation through the Freigeist Program. G.G. and R.P. are grateful to the Ministero dell'Università e della Ricerca (MUR) and the University of Pavia for funding through the "Dipartimenti di Eccellenza 2023–2027" programme and the FARE (Framework per l'Attrazione e il Rafforzamento delle Eccellenze per la ricerca in Italia) project EXPRESS ("Exploring Photoferroelectricity in Halide Perovskites for Optoelectronics"), under the project "Development and Characterization of Halide Perovskites for Photoferroelectric Applications". G.G. acknowledges funding from the Fondazione Cariplo Economia Circolare 2021 project "Green Flexible Hybrid Perovskite Solar Module for the Market: From Smart Lead Manipulation to Recycling" (FLHYPER, No. 20201067), funded under the "Circular Economy–2020" call. G.G. also acknowledges the GOPV project (CSEAA_00011), which received funding through the Bando Ricerca di Sistema—CSEA—TIPO A, Piano Triennale 2019–2021, under Decreto Direttoriale of 27 October 2021 of the Ministero della Transizione Ecologica. F.F.-S. acknowledges funding from the Ministerio de Ciencia, Innovacion y Universidades through project PID2023-149158OB-C41. The authors thank Clara A. Aranda for helpful discussions.

**ORCID**

Agustin O. Alvarez: 0000-0002-0920-5390;

Biruk Alebachew Seid: 0009-0009-1123-3866

Riccardo Pallotta: 0009-0007-6033-8280;

Elena Mas-Marzá: 0000-0002-2308-0635

Giulia Grancini: 0000-0001-8704-4222;

Felix Lang: 0000-0001-9711-380X

Francisco Fabregat-Santiago: 0000-0002-7503-1245

Bruno Ehrler: 0000-0002-5307-3241

# Supporting information

## Recombination or Transport? Understanding Electronic and Ionic Processes in Perovskite Solar Cells by Combining Impedance and Light-Modulation Spectroscopies

Agustin O. Alvarez,[a]* Athira Shaji,[a] Max Mundt,[a] Biruk A. Seid,[b] Riccardo Pallotta,[c] Elena Mas-Marzá,[d] Giulia Grancini,[c] Felix Lang,[b] Francisco Fabregat-Santiago,[d]* and Bruno Ehrler[a]*

[a] LMPV - Sustainable Energy Materials Department, AMOLF, 1098 XG Amsterdam, The Netherland

[b] ROSI Freigeist Juniorgroup, Institut für Physik und Astronomie, University of Potsdam, 14476 PotsdamGolm, Germany

[c] Department of Chemistry, University of Pavia, Via T. Taramelli 14, 27100 Pavia, Italy

[d] Institute of Advanced Materials, Universitat Jaume I, 12006 Castelló de la Plana, Spain

E-mail: a.alvarez@amolf.nl; fabresan@uji.es; b.ehrler@amolf.nl

## S1. Equivalent Circuit Models

The equivalent circuits in **Figure 1**g and **Figure 1**h reproduce the IS arc in **Figure 1**b through the frequency dependence of the capacitor impedance, which tends to zero at high frequencies and to infinity at low frequencies. At high frequencies, the capacitor charges and discharges without building up an appreciable voltage, so the resistance in parallel with it ($R_{rec}$ or $R_{tr}$) is not seen. The high-frequency limit of $Z$ differs between the circuits. It is $R_S$ in the recombination circuit, whereas in the transport circuit it is $R_S + R_{rec}$, because with $j_{ph}$ open, the modulated current between the terminals must still pass through $R_{rec}$. The two circuits therefore give the same IS spectrum if $R_S$ in the recombination circuit is replaced by $R_S + R_{rec}$ in the transport circuit, and IS cannot tell them apart. IMVS and IMPS, which modulate $j_{ph}$, can. In IMVS, no current is extracted, so in the recombination circuit (**Figure 1**g) the modulated $j_{ph}$ flows only through the RC element. At high frequencies, it charges and discharges the capacitor without building up an appreciable voltage, so $W$ tends to zero. At low frequencies it flows through $R_{rec}$ instead, so $W$ approaches a real value set by $j_{ph}$ and $R_{rec}$, tracing an arc. In IMPS, the RC element competes with the extraction path through $R_S$. At high frequencies, the capacitor again takes up the modulated charge, so $Q$ tends to zero, whereas at low frequencies $Q$ approaches the fraction of $j_{ph}$ that flows through $R_S$, competing with $R_{rec}$. In the transport circuit (**Figure 1**h), the extraction path ($R_S$ plus the RC) carries no current in IMVS, so $W$ is set only by $j_{ph}$ and $R_{rec}$ and shows no arc. In IMPS, the extraction path is $R_S$ at high frequencies, where the capacitor charges and discharges freely and $R_{tr}$ is not seen, and $R_S$ plus $R_{tr}$ at low frequencies. Because this path competes with $R_{rec}$, $Q$ decreases from a positive value as frequency decreases, reproducing the inverted arc.

For the simulations in **Figure 1**, the recombination equivalent circuit (**Figure 1**g) used $R_S = 2.5\,\Omega\,\mathrm{cm}^2$, $R_{\mathrm{rec}} = 10\,\Omega\,\mathrm{cm}^2$ and $C_{\mathrm{rec}} = 1\,\mu\mathrm{F\,cm}^{-2}$, while the transport equivalent circuit (**Figure 1**h) used $R_S = 0.5\,\Omega\,\mathrm{cm}^2$, $R_{\mathrm{rec}} = 2\,\Omega\,\mathrm{cm}^2$, $R_{\mathrm{tr}} = 10\,\Omega\,\mathrm{cm}^2$ and $C_{\mathrm{tr}} = 1\,\mu\mathrm{F\,cm}^{-2}$. In both cases, the product of absorptance and charge-separation efficiency was $a\eta_{\mathrm{sep}} = 0.8$.

To represent and quantify the responses in **Figure 2**b-d, we use the equivalent circuit in **Figure 2**e. It consists of a series resistance $R_S$ and two branches in parallel with $j_{ph}$: a recombination resistance $R_{rec}$ (blue) in parallel with a capacitor $C_{rec}$, which can be chemical or geometric as discussed in Section 2.1, and a branch (green) formed by $R_{ion}$ in series with a constant phase element ($CPE$), which accounts for the depressed shape of the low-frequency arc [1-3]. The spectra follow from the frequency dependence of $C_{rec}$ and the $CPE$, whose impedances are small at high frequencies and large at low frequencies. At the highest frequencies, $C_{rec}$ takes up the modulated current by charging and discharging without building up an appreciable internal voltage, so the impedance is $R_S$ and $W$ and $Q$ tend to zero. Since no transport process is resolved, $R_S$ includes any transport resistance. As frequency decreases and $C_{rec}$ fills, the modulated charge sees the parallel combination of $R_{rec}$ and $R_{ion}$, producing the high-frequency arc in the three techniques. At lower frequencies, the $CPE$ progressively blocks current through $R_{ion}$, leaving $R_{rec}$ as the only internal branch and producing the second arc in all three techniques. This behavior can represent ionic charge accumulation coupled to recombination, with the AC charging current decreasing toward low frequencies, analogous to the charge-storage contribution to the high-frequency response. The same branch can also describe ion-mediated suppression of recombination. In this interpretation, it represents an additional effective recombination contribution at frequencies above the ionic relaxation that is progressively suppressed toward lower frequencies. The CPE then represents an apparent capacitive response associated with delayed recombination modulation and need not correspond to an equivalent amount of stored charge. Both interpretations reproduce the observed increases in resistance, photovoltage, and photocurrent, and both contributions may coexist.

To analyze and quantify the responses in **Figure 3**b-d, we use the equivalent circuit in **Figure 3**e, which is the equivalent circuit of **Figure 2**e with an additional RC pair, $R_{tr}$ in parallel with a capacitor $C_{tr}$ (red), placed in series with $j_{ph}$ and the terminals. This combines the two cases of **Figure 1**g and **Figure 1**h, with transport in series and recombination in parallel with $j_{ph}$. The frequency response follows the same logic as in **Figure 2**. At the highest frequencies, the capacitors take up the modulated charge without building up an appreciable voltage, so $Z$ is $R_S$ and $W$ and $Q$ tend to zero. As the frequency decreases and $C_{rec}$ fills, the resistive pathways become visible, producing the high-frequency arc in the three techniques. At this stage $C_{tr}$ still takes up the modulated charge, so the extraction path is $R_S$ alone. At lower frequencies $C_{tr}$ fills and $R_{tr}$ is added to the extraction path, which produces the mid-frequency arc in IS and lowers the extracted current in IMPS, giving the inverted arc. IMVS is unaffected because no current is extracted and none flows through the series path. Finally, at the lowest frequencies, the $CPE$ blocks the current through $R_{ion}$, leaving $R_{rec}$ as the only internal branch and producing the low-frequency arc in IS, IMVS, and IMPS, as in **Figure 2**.

To analyze and quantify the responses in **Figure 4**b-d, we use the equivalent circuit in in **Figure 4**e, which is the equivalent circuit of **Figure 2**e with an inductor $L$ replacing the $CPE$ in the low-frequency branch ($R_{ion}$ in series with $L$, green). The frequency response follows the same logic as before, now taking into account that the inductor behaves as an open circuit at high frequencies and as a short circuit at low frequencies. At the highest frequencies, $C_{rec}$ takes up the modulated charge without building up an appreciable voltage, so $Z$ is $R_S$ and $W$ and $Q$ tend to zero. As the frequency decreases and $C_{rec}$ fills, recombination through $R_{rec}$ produces the high-frequency arc (blue) in the three techniques, with the additional channel still blocked. At lower frequencies, the inductor allows current to flow through $R_{ion}$, opening this additional recombination channel in parallel with $R_{rec}$ and lowering the total recombination resistance. This produces the inverted arc in IS and IMVS, and it also reduces the extracted photocurrent in IMPS, although this effect is small in the measured spectrum.

## S2. Fitting Parameters

**Table S1** Parameters obtained from the global fitting of the IS, IMPS, and IMVS spectra of the two-arc, three-arc, and inverted-arc devices in **Figure 2**, **Figure 3**, and **Figure 4**, with the equivalent circuits in **Figure 2**e, **Figure 3**e, and **Figure 4**e, respectively.

| | **High-Efficiency PSC: Two-Arc Impedance** | **Non-Optimized Contacts PSC: Three-Arc Impedance** | **Lower Efficiency PSC: Inverted-Arc Impedance** |
|---|---|---|---|
| | **Figure 2** | **Figure 3** | **Figure 4** |
| | 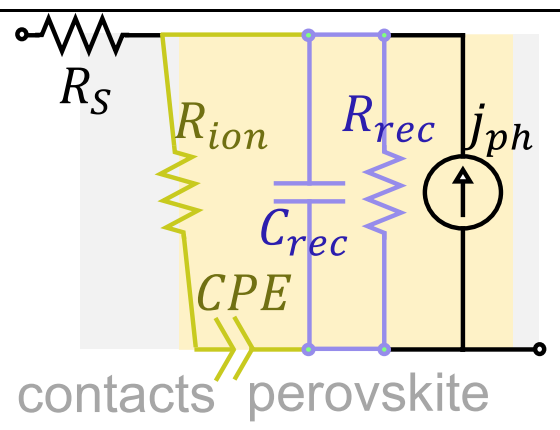 | 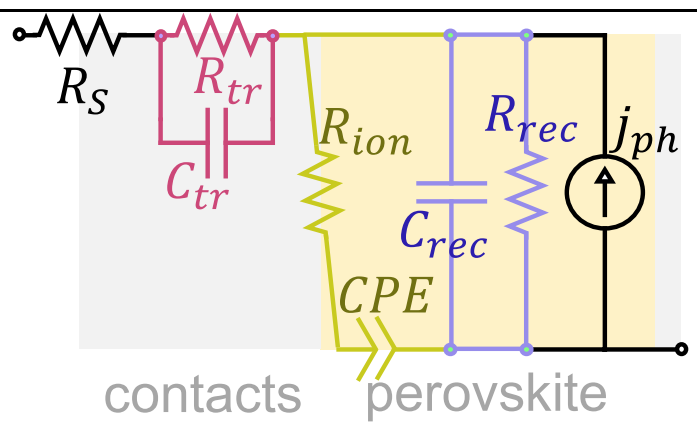 | 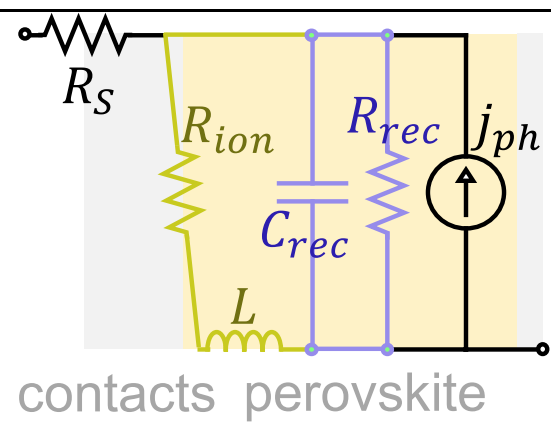 |
| $R_S$ [Ω cm$^2$] | 5.47 ± 0.09 | 17.27 ± 0.13 | 5.42 ± 0.08 |
| $R_{rec}$ [Ω cm$^2$] | 14.0 ± 0.2 | 8.2 ± 0.2 | 5.7 ± 0.3 |
| $C_{rec}$ [nF cm$^{-2}$] | 170 ± 1 | 141 ± 3 | 116.3 ± 1.2 |
| $R_{ion}$ [Ω cm$^2$] | 22.3 ± 0.6 | 19.3 ± 0.8 | 90 ± 5 |
| $\alpha$ [ ] | 0.58 ± 0.02 | 0.47 ± 0.02 | |
| $CPE$ [mF s$^{\alpha-1}$ cm$^{-2}$] | 11.4 ± 1.2 | 8 ± 3 | |
| $L$ [mH cm$^2$] | | | 55 ± 7 |
| $R_{tr}$ [Ω cm$^2$] | | 30.7 ± 0.1 | |
| $C_{tr}$ [nF cm$^{-2}$] | | 431 ± 4 | |
| $\tau_{rec}$ [$\mu$s] | 1.46 ± 0.02 | 0.81 ± 0.02 | 0.66 ± 0.04 |
| $\tau_{tr}$ [$\mu$s] | | 13.24 ± 0.12 | |
| $\tau_{ion}$ [ms] | 216 ± 41 | 36 ± 33 | 0.57 ± 0.08 |
| $a$ [%] | 96.00 ± 1.00 | 96.00 ± 1.00 | 93.00 ± 1.00 |
| $\eta_{sep}$ [%] | 74.0 ± 0.9 | 34.7 ± 0.8 | 78.4 ± 1.0 |
| $\eta_{col}$ [%] | 71.9 ± 0.5 | 14.6 ± 0.3 | 49.5 ± 1.3 |
| $IQE_{PV\text{-}Diff}$ [%] | 53.2 ± 0.7 | 5.1 ± 0.2 | 38.8 ± 1.1 |
| $EQE_{PV\text{-}Diff}$ [%] | 51.0 ± 0.4 | 4.9 ± 0.1 | 36.1 ± 1.0 |

The series resistance ($R_S$), the recombination resistance ($R_{rec}$) and capacitance ($C_{rec}$), and, for each device, the elements of the other processes ($R_{ion}$ and the $CPE$ with exponent α in the two-arc and three-arc devices, $R_{ion}$ and $L$ in the inverted-arc device, and $R_{tr}$ and $C_{tr}$ in the three-arc device) were obtained directly from the fits of each set of measurements. From these parameters, the characteristic time of each process ($\tau_{rec}$, $\tau_{tr}$, and $\tau_{ion}$) was calculated, which is comparable to the inverse of the frequency at the extremum of the corresponding arc in IS and IMVS. The fits also gave the product of the absorptance ($a$) and the charge-separation efficiency ($\eta_{sep}$), $a \cdot \eta_{sep}$. Combined with the absorptance measured by UV-vis, it gave $\eta_{sep}$. From these values, the charge-collection efficiency ($\eta_{col}$) and the differential internal and external quantum efficiencies ($IQE_{PV\text{-}Diff}$ and $EQE_{PV\text{-}Diff}$) were calculated following the equations in [4].